\documentclass[conference]{IEEEtran}

\usepackage{amsmath,amssymb}
\usepackage{graphicx}
\usepackage{booktabs}
\usepackage{multirow}
\usepackage{url}
\usepackage[hidelinks]{hyperref}

\title{SIDSense: Database-Free TV White Space Sensing for Disaster-Resilient Connectivity}

\author{%
  \IEEEauthorblockN{George M. Gichuru}
  \IEEEauthorblockA{Amini, Nairobi, Kenya\\
  george@amini.ai}
  \and
  \IEEEauthorblockN{Zoe Aiyanna M. Cayetano}
  \IEEEauthorblockA{Amini, Nairobi, Kenya\\
  zoe@amini.ai}
}

\begin{document}
\maketitle

\begin{abstract}
Small Island Developing States (SIDS) are disproportionately exposed to climate-driven disasters, yet often rely on fragile terrestrial networks that fail when they are most needed. TV White Space (TVWS) links offer long-range, low-power coverage; however, current deployments depend on Protocol to Access White Spaces (PAWS) database connectivity for channel authorization, creating a single point of failure during outages.

We present \textit{SIDSense}, an edge AI framework for database-free TVWS operation that preserves regulatory intent through a compliance-gated controller, audit logging, and graceful degradation. SIDSense couples CNN-based spectrum classification with a hybrid ``sensing-first, authorization-as-soon-as-possible'' workflow and co-locates sensing and video enhancement with a private 5G stack on a maritime vessel to sustain situational-awareness video backhaul.

Field experiments in Barbados demonstrate sustained connectivity during simulated PAWS outages, achieving 94.2\% sensing accuracy over 470--698~MHz with 23~ms mean decision latency, while maintaining zero missed 5G Layer-1 (L1) deadlines under GPU-aware scheduling. We release an empirical Caribbean TVWS propagation and occupancy dataset and look to contribute some of the components of the SIDSense pipeline to the open source community to accelerate resilient connectivity deployments in climate-vulnerable regions.
\end{abstract}


\section{Introduction}

Small Island Developing States occupy a unique position in the global climate crisis: geographically exposed to tropical cyclones, tsunamis, and sea-level rise while simultaneously constrained by limited telecommunications infrastructure and economic resources for rapid recovery. The United Nations Office for Disaster Risk Reduction reports that SIDS experience disaster-related mortality rates exceeding double the global average, with average annual GDP losses of 2.1\% compared to 0.3\% elsewhere~\cite{ref:undrr2022}. When Hurricane Maria devastated Dominica in 2017, the island lost 90\% of its cellular infrastructure within hours, leaving emergency responders without communication pathways precisely when coordination was most critical~\cite{ref:itu2017dominica}.

TV White Space (TVWS) technology---exploiting unused UHF television spectrum between 470--698~MHz---offers compelling advantages for disaster-resilient connectivity in island environments. The favorable propagation characteristics of sub-GHz frequencies enable long-range over-water links exceeding 20~kilometers, while the relatively low power requirements align with solar-powered emergency deployments~\cite{ref:mwangama2015africon}. Microsoft's pioneering deployments in the Philippines following Typhoon Haiyan (2013) and in Puerto Rico after Hurricane Maria (2017) demonstrated TVWS's viability for post-disaster connectivity restoration~\cite{ref:microsoft2019airband}.

However, current TVWS systems depend fundamentally on geolocation database access for channel authorization. The Protocol to Access White Spaces (PAWS), specified in RFC~7545~\cite{ref:rfc7545}, requires devices to query a whitespace database (WSDB) before transmitting. This creates a disaster-critical failure mode: when the backhaul used to reach the WSDB is degraded or unavailable, TVWS equipment may be forced silent even if clean spectrum is locally observable.

The challenge is compounded by regulatory absence in many SIDS regions. Examination of the IANA PAWS Ruleset ID Registry reveals no registered rulesets for Caribbean nations including Barbados, Jamaica, Trinidad and Tobago, or the Eastern Caribbean states~\cite{ref:iana_paws}. Pacific island nations similarly lack PAWS infrastructure. This regulatory vacuum means that even when database connectivity exists, the query mechanisms may return null authorizations for SIDS deployments.

We address these challenges through \textit{SIDSense}, an edge AI framework that enables compliant TVWS operation when PAWS database access is degraded or unavailable. Our key insight is to treat spectrum sensing not as a replacement for database governance, but as a resilience mechanism: sensing produces probabilistic channel recommendations that (i) are cross-checked against WSDB grants when connectivity exists and (ii) can support time-bounded emergency operation under pre-cleared waivers, with complete audit trails for post-event review.

\subsection{Contributions}
The contributions of this paper are fourfold:
\begin{enumerate}
  \item We present the first database-free ML spectrum sensing framework specifically designed for TVWS bands in SIDS disaster scenarios, addressing the previously unexplored intersection of cognitive radio research and island nation resilience requirements.
  \item We develop a compliance-gated mode controller that jointly optimizes video quality and spectrum efficiency while maintaining full audit trails suitable for regulatory review. This enables a sensing-first workflow with authorization as soon as connectivity is restored, as required for emergency exemptions.
  \item We demonstrate GPU-aware RAN scheduling that co-locates spectrum sensing and super-resolution AI workloads with 5G L1/PHY processing while maintaining zero missed symbol deadlines, advancing the practical deployment of AI-on-RAN architectures in resource-constrained environments.
  \item We contribute the first empirical TVWS propagation measurements and spectrum occupancy dataset for Caribbean island environments, validated through field deployments in Barbados with planned hurricane drill integration during CaribeWave~2026.
\end{enumerate}

The remainder of this paper is organized as follows. Section~\ref{sec:related} surveys related work in TVWS systems, spectrum sensing, and disaster communications. Section~\ref{sec:model} formalizes the problem and introduces our system model. Section~\ref{sec:architecture} details the SIDSense architecture and its component algorithms. Section~\ref{sec:evaluation} presents experimental methodology and results from Barbados deployments. Section~\ref{sec:discussion} discusses regulatory implications and deployment considerations. Section~\ref{sec:conclusion} concludes with directions for future work.

\section{Related Work}\label{sec:related}
\subsection{TV White Space Systems and Database Governance}
The regulatory framework for TVWS evolved from the FCC's 2008 ruling permitting unlicensed operation in television bands, contingent on geolocation database consultation~\cite{ref:fcc2008}. The PAWS protocol, standardized by IETF in 2015, defines the query-response mechanism through which devices obtain channel availability and maximum permissible power levels~\cite{ref:rfc7545}. Commercial TVWS databases operated by Google, Microsoft, and others have enabled deployments across North America, Europe, and parts of Africa and Asia~\cite{ref:kawade2010dyspan}.

Academic research on TVWS has concentrated on three areas: propagation modeling for the UHF band~\cite{ref:stevenson80222}, interference protection for incumbent television services~\cite{ref:nekovee2010survey,ref:ghosh2012broadcast}, and network protocols for secondary access~\cite{ref:nekovee2010survey}. Foundational work by Murty \textit{et al.} demonstrated practical TVWS operation but assumed reliable database connectivity~\cite{ref:murty2012senseless}. Subsequent deployment studies in Kenya, South Africa, and the Philippines similarly operated within the database-governed paradigm~\cite{ref:zennaro2013africon,ref:lysko2013africon,ref:falade2013commag}.

The limitation of existing TVWS research for SIDS contexts is the implicit assumption of stable internet backhaul for database queries. Johnson and van Stam's critique of Western-centric protocol assumptions in Global South deployments applies directly: PAWS was designed for environments where connectivity interruptions are exceptional rather than routine, and where regulatory infrastructure exists to populate database records~\cite{ref:johnson2017dev}.

\subsection{Spectrum Sensing Techniques}
Spectrum sensing research predates the database-centric approach and was originally positioned as the primary mechanism for cognitive radio systems to detect incumbent transmissions~\cite{ref:haykin2005cr}. Y\"ucek and Arslan's survey categorizes techniques into energy detection, matched filtering, cyclostationary feature detection, and more recently, machine learning-based classification~\cite{ref:yucek2009survey}.

Deep learning approaches have demonstrated significant advances in spectrum sensing performance. Zheng \textit{et al.} achieved 95.3\% detection accuracy using convolutional neural networks on spectrogram representations~\cite{ref:zheng2020chinacom}. O'Shea \textit{et al.} introduced the RadioML dataset and demonstrated that CNNs could learn modulation classification directly from raw IQ samples~\cite{ref:oshea2018jsps}. Riyaz \textit{et al.} demonstrated CNN-based radio identification as a practical building block for RF classification pipelines~\cite{ref:riyaz2018commag}. Restuccia \textit{et al.} applied transfer learning to reduce training requirements for new deployment environments~\cite{ref:restuccia2019mobihoc}.

However, existing ML spectrum sensing research has not addressed TVWS-specific requirements. The UHF television band presents unique characteristics including narrowband incumbent signals with known spectral signatures, relatively static channel assignments, and regulatory constraints on detection probability. Our work adapts CNN-based sensing specifically for the 470--698~MHz band and integrates sensing decisions with the compliance requirements of TVWS operation.

\subsection{Disaster Communications and Network Resilience}
Disaster communications research has examined rapidly deployable networks~\cite{ref:bao2010commag}, mesh architectures~\cite{ref:liu2013icist}, and satellite-terrestrial integration~\cite{ref:saarnisaari2019aess}. Pratas \textit{et al.} proposed cooperative spectrum sensing for disaster relief scenarios, demonstrating that distributed sensing can improve detection reliability when individual nodes experience fading~\cite{ref:pratas2012vtc}. Their work focused on LTE bands rather than TVWS and did not address the database connectivity problem.

Maritime communications present additional challenges relevant to SIDS deployments. The two-ray propagation model dominates over-water paths beyond the breakpoint distance, creating periodic fading patterns that complicate link budget analysis~\cite{ref:iturp1546}. Recent work on maritime 5G networks has addressed these propagation effects but not TVWS applications~\cite{ref:wang2021iotj}.

\subsection{AI-RAN and Edge Computing Architectures}
The co-location of AI inference with RAN functions has emerged as a significant research direction. Polese \textit{et al.} survey Open RAN opportunities for intelligent network control through xApps and rApps in the O-RAN architecture~\cite{ref:polese2023oran}. The AI-RAN Alliance, established in 2024 with participation from industry and academic partners, has defined reference architectures for GPU-accelerated base stations~\cite{ref:airan2024report}.

Bonati \textit{et al.}'s ColO-RAN framework demonstrated practical co-location of ML inference with RAN functions, achieving real-time control loop latencies suitable for scheduling decisions~\cite{ref:bonati2023coloran}. The cuSense project showed that GPU-accelerated sensing could operate alongside RAN L1 processing with appropriate resource isolation~\cite{ref:tehrani2023cusense}.

\section{System Model and Problem Formulation}\label{sec:model}
\subsection{Deployment Scenario}
We consider a maritime vessel operating in coastal waters of a Small Island Developing State during or immediately following a disaster event. The vessel serves as a mobile communications hub providing situational awareness capabilities for emergency response. An unmanned aerial vehicle (UAV) equipped with a 360-degree camera operates from the vessel, capturing real-time video of coastal infrastructure, flood extent, and potential survivors.

The vessel hosts a private 5G network comprising a complete 5G core (AMF/SMF/UPF), a gNB with centralized unit (CU), distributed unit (DU), and an O-RAN 7.2x compliant radio unit (O-RU) operating in a licensed or trial FR1 band. The UAV connects as a 5G user equipment, transmitting video over the air interface to the vessel-based edge compute platform.

Backhaul connectivity to shore-based emergency operations centers is provided through a TVWS link operating in the 470--698~MHz band. This frequency range offers favorable propagation for the 10--30~km over-water paths typical of island coastal operations. The TVWS equipment must query a PAWS-compliant database for channel authorization; however, the terrestrial infrastructure supporting database connectivity may be damaged or congested during disaster conditions.

\subsection{Channel Model}
The TVWS propagation environment over water exhibits characteristics well-described by the two-ray model beyond the breakpoint distance. For transmitter height $h_t$ and receiver height $h_r$, the breakpoint distance is
\begin{equation}
  d_b = \frac{4 h_t h_r}{\lambda},
\end{equation}
where $\lambda$ is the wavelength. For typical TVWS deployments with shore mast at 25~m, vessel antenna at 5~m, and center frequency 550~MHz, the breakpoint occurs at approximately 1.8~km. Beyond this distance, path loss follows a steeper slope due to destructive interference between direct and sea-surface reflected rays.

We model received signal power at distance $d$ as
\begin{equation}
  P_r(d) = P_t + G_t + G_r - L(d) - M_f,
\end{equation}
where $P_t$ is transmit power (limited by PAWS grants, typically 36~dBm EIRP for mobile devices), $G_t$ and $G_r$ are antenna gains, $L(d)$ is distance-dependent path loss following ITU-R P.1546 for mixed land-sea paths, and $M_f$ is a fade margin derived from route-measured statistics to achieve 99\% link availability.

\subsection{Spectrum Sensing Model}
The spectrum sensing subsystem receives wideband IQ samples covering the 470--698~MHz TVWS allocation. Let $x(n)$ denote the received signal at discrete time $n$. The sensing problem is formulated as a binary hypothesis test for each 6/7/8~MHz channel:
\begin{align}
  H_0 &: x(n) = w(n) && \text{(channel idle)}\\
  H_1 &: x(n) = s(n) + w(n) && \text{(incumbent present)}
\end{align}
where $s(n)$ represents incumbent television transmissions and $w(n)$ is additive white Gaussian noise. Unlike traditional spectrum sensing that aims for simple detection, we require channel classification that distinguishes between: (a) active television broadcast, (b) wireless microphone operation, (c) other TVWS secondary users, and (d) vacant spectrum.

\subsection{Problem Statement}
Our objective is to maximize the delivery quality of situational awareness video from the UAV to shore-based emergency operations, subject to: (i) spectrum compliance (valid PAWS grant or emergency waiver with audit logging), (ii) RAN timing (zero missed L1/PHY deadlines under co-tenancy), (iii) energy constraints, and (iv) graceful degradation of the video stream via super-resolution when the UAV--vessel 5G link (uplink and downlink) is degraded.

\section{SIDSense Architecture}\label{sec:architecture}
\subsection{System Overview}
\begin{figure*}[t]
  \centering
  \includegraphics[width=\textwidth]{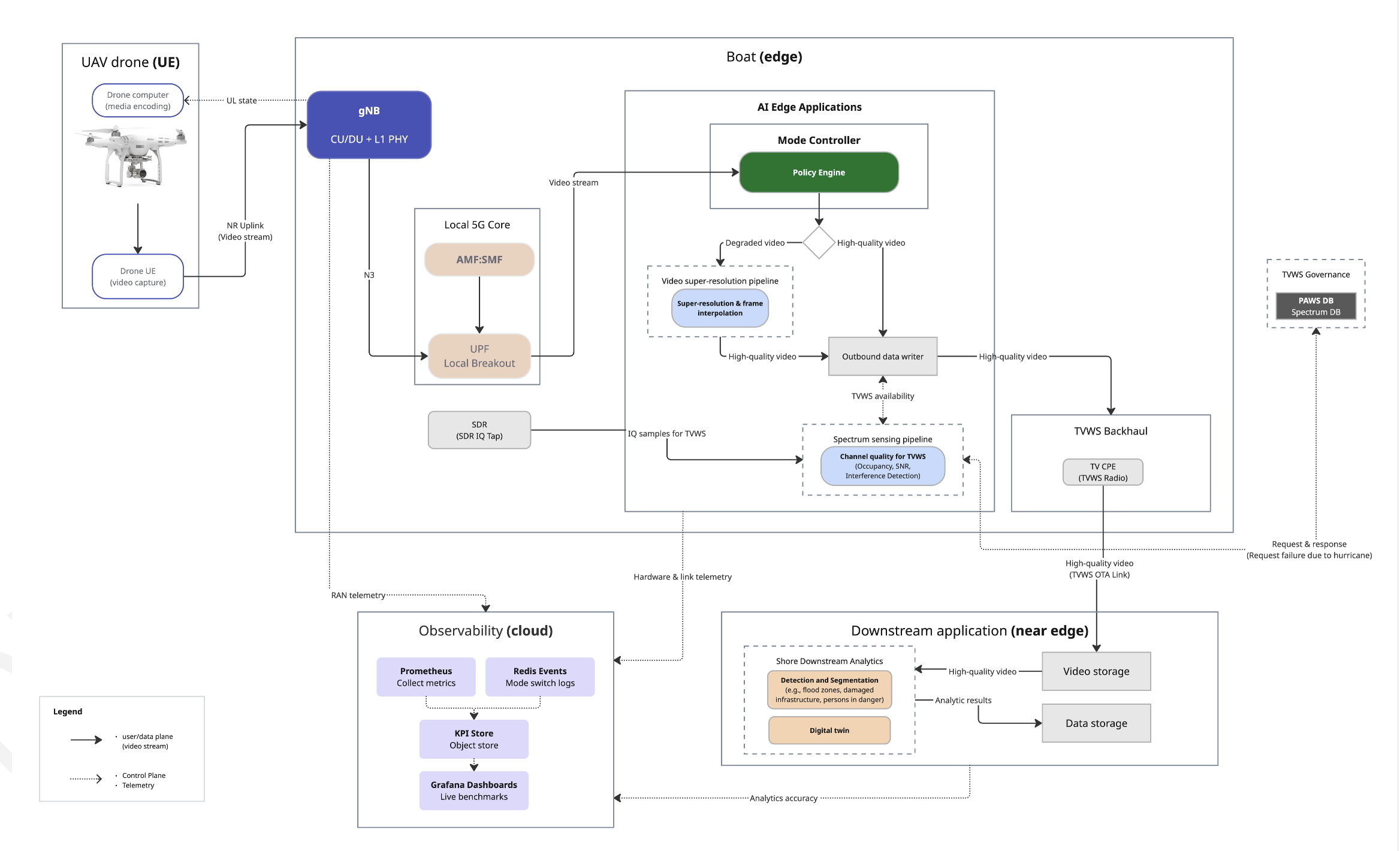}
  \caption{SIDSense reference architecture: a UAV video source uplinks to a vessel-hosted private 5G network with UPF local breakout and edge AI services; a compliance gate authorizes TVWS backhaul based on PAWS grants or sensing-first emergency policy; all decisions are audit-logged and exported to shore when connectivity permits.}
  \label{fig:reference-arch}
\end{figure*}

Figure~\ref{fig:reference-arch} illustrates the end-to-end SIDSense architecture, organized as four integrated planes: the UAV source (left), the vessel-based RAN and edge compute (center), the TVWS backhaul and governance (right), and the observability and downstream analytics (bottom/shore). Within these planes, SIDSense comprises four integrated subsystems operating on a unified edge compute platform: (1) a spectrum sensing pipeline that provides channel recommendations, (2) a super-resolution video processing pipeline that compensates for bandwidth constraints, (3) a compliance-gated mode controller that coordinates between sensing, video processing, and RAN determinism, and (4) a digital twin that provides propagation priors and regulatory-ready audit trails.

The vessel edge platform hosts the complete 5G stack (5GC and gNB CU/DU) alongside AI inference workloads. We adopt a GPU-aware scheduling approach where the DU/L1 pipeline executes on dedicated CUDA streams with highest priority, while spectrum sensing and video enhancement execute on separate partitions. CPU cores are isolated between RAN control plane functions (pinned to specific cores) and AI preprocessing.

Video captured by the UAV traverses the 5G air interface to the vessel gNB, where UPF local breakout steers it either to the super-resolution (SR) pipeline (if operating in degraded mode) or directly to the outbound data writer. The spectrum sensing pipeline (lower center) continuously monitors TVWS availability, feeding channel recommendations to the compliance gate that governs backhaul transmission. A mode controller xApp (upper center) orchestrates UAV video encoding and SR activation based on sustained 5G uplink conditions and GPU resource availability.

SIDSense comprises two categories of AI workloads with distinct architectural roles. AI-for-networking functions operate in the control plane (spectrum sensing recommendations and mode control), while AI-for-media functions operate in the user plane (super-resolution and frame interpolation). The coupling point between these planes is the mode tag: control-plane decisions set the tag, which determines whether user-plane processing is enabled (SR) or bypassed.

\subsection{CNN-Based Spectrum Sensing Pipeline}
Spectrum sensing follows a simple, reproducible abstraction: \textit{RF capture $\rightarrow$ representation extraction $\rightarrow$ classifier $\rightarrow$ decision}. The pipeline ingests wideband IQ samples from a dedicated software-defined radio (SDR) frontend covering 470--698~MHz. For field deployments we use an Ettus USRP N310 with a GPS-disciplined reference; the receiver is configured for wideband capture (up to 100~MHz instantaneous bandwidth), requiring multiple sequential captures to cover the full TVWS allocation. For community replication and prototyping, a USRP B210 with an external GPSDO may be substituted with reduced scan cadence.

Regardless of SDR choice, real TVWS environments can include strong adjacent incumbents; we therefore use an external 470--698~MHz bandpass filter (and, when needed, a low-noise amplifier (LNA)) to improve out-of-band rejection and effective dynamic range.

Representation extraction uses short-time Fourier transform with a 1024-point FFT, 50\% overlap, and a Hanning window, producing $224\times224$ log-magnitude spectrogram tensors for each channel-time window.

We implement the classifier as a ResNet18 backbone pretrained on ImageNet with the final fully-connected layer replaced by a 256-unit dense layer and a 4-class softmax output (TV broadcast, wireless microphone, other TVWS, vacant). Transfer learning provides robust feature extraction despite limited Caribbean training data.

The model is trained on a combination of over-the-air captures from Barbados and synthetic TVWS waveforms. We employ data augmentation including frequency shifting ($\pm 500$~kHz), time stretching ($0.9\times$--$1.1\times$), and additive noise injection.

For each channel, the sensing pipeline outputs $(\text{class}, \text{confidence})$ where $\text{confidence}\in[0,1]$ is the softmax probability. Channels classified as vacant with confidence exceeding threshold $\theta_{\text{sense}}=0.85$ are considered candidates for TVWS transmission; for auditability, we additionally record a binary occupancy verdict (occupied vs.
 vacant) derived from the multi-class posteriors.

\subsection{Super-Resolution Video Pipeline}
When sustained 5G uplink degradation prevents reliable transmission of high-definition video from the UAV, the system switches to degraded mode: the UAV transmits a lower-resolution stream (480p at 15~fps, approximately 0.8--1.2~Mbps after H.265 encoding) over the 5G air interface. The super-resolution pipeline running on the vessel edge platform upscales this to perceptual 1080p at 30~fps.

We employ a two-stage approach combining spatial upscaling via Real-ESRGAN~\cite{ref:realesrgan2021} and temporal interpolation via RIFE~\cite{ref:rife2022}. The pipeline stages are: decode (NVDEC), spatial upscale (Real-ESRGAN-x4plus), temporal interpolation (RIFE, 15\,fps $\rightarrow$ 30\,fps), and output encode (NVENC) or raw streaming to analytics.

\subsection{UPF Local Breakout and End-to-End Video Path}
Video frames captured by the UAV are encoded by the companion computer according to the current mode (1080p/30~fps H.265 for \textit{Native-HD}, 480p/15~fps for \textit{Degraded}) and transmitted over the 5G uplink. The gNB forwards packets to the UPF, which performs local breakout at the vessel edge rather than backhauling to a remote core. Traffic steering is mode-aware: \textit{Degraded}-mode streams are routed to the SR pipeline before reaching the outbound data writer, while \textit{Native-HD} streams bypass enhancement to minimize latency.

The outbound data writer queues frames for TVWS backhaul transmission. The compliance gate verifies that a valid PAWS grant exists for the selected channel, or---during connectivity outages---that an emergency waiver is active and sensing confidence exceeds $\theta_{\text{sense}}=0.85$. All transmission decisions are logged with timestamp, GPS coordinates, channel, EIRP, grant status, and sensing confidence.

System observability is derived from both RAN and edge telemetry. The xApp consumes RAN-side KPMs (uplink throughput, PRB utilization, CQI/BLER) together with GPU utilization and thermal headroom; the sensing service records per-channel class posteriors and decision latency; and the backhaul subsystem records link-layer goodput and packet loss. At the shore station, the downstream analytics pipeline performs object detection and segmentation on received video, feeding results to a digital twin for emergency operations center visualization.

\subsection{Compliance-Gated Mode Controller}
The mode controller is implemented as a Near-RT RIC xApp that subscribes to RAN key performance measurements (KPMs) and edge platform telemetry. It implements a hysteresis-based state machine with asymmetric thresholds and minimum dwell times to prevent mode flapping while responding to sustained degradation within seconds.

Two primary operating modes are supported: \textit{Native-HD} (UAV transmits 1080p directly; SR bypassed) and \textit{Degraded} (UAV transmits 480p/15~fps; SR active). Mode transitions are driven by sustained 5G uplink conditions (e.g., uplink throughput, PRB utilization, CQI/BLER) and GPU resource availability (utilization and thermal headroom) to ensure user-plane AI workloads never compromise RAN L1 timing determinism. The xApp signals the UAV companion computer to switch encoder settings (native to degraded) and enables or disables SR accordingly.

Backhaul transmission remains governed by TVWS compliance: channel changes and emission decisions occur only under a valid PAWS grant, or during outages under a pre-cleared emergency waiver with sensing confidence exceeding $\theta_{\text{sense}}=0.85$.

Policy configuration (threshold profiles and emergency-mode parameters) may be managed by a Non-RT RIC rApp; however, rApp integration is outside the scope of the current implementation.

\subsection{Digital Twin and Audit Logging}
A digital twin maintains a probabilistic model of the RF environment, incorporating coastline geometry, licensed TV transmitter locations and contours, historical occupancy patterns, and optional weather/sea-state data. Every transmission decision is logged with timestamp, GPS coordinates, selected channel, EIRP, grant validity status, and sensing confidence for regulatory-ready post-event review. Logs are cryptographically signed and stored locally and transmitted to shore when backhaul permits.

\section{Experimental Evaluation}\label{sec:evaluation}

\subsection{Testbed Configuration}
We deployed the SIDSense testbed in Barbados through collaboration with the Barbados Telecommunications Unit. The shore station is located at a coastal site with antenna height 25~m above sea level. The vessel platform operates a 12~m research vessel with antenna height 5~m.

Hardware configuration comprises: edge server (Supermicro SuperServer SYS-222HE-FTN, dual Intel GNR-SP 6760P 64 cores/128 threads, 512GB RAM, 1 NVIDIA H200 NVL for L1/PHY and AI workloads with NVIDIA Multi-instance GPU sharding), SDR (Ettus USRP N310, GPS-disciplined reference, custom 470--698~MHz bandpass filter), 5G RAN (OpenAirInterface gNB with NVIDIA Aerial cuPHY and O-RU in n78 trial license), TVWS radios (Adaptrum ACRS 2.0 certified devices, 6~MHz channels, max 36~dBm EIRP), and UAV (DJI M300 RTK with 360\textdegree{} camera, 5G modem, Jetson companion computer).

\subsection{Spectrum Sensing Performance}
Overall accuracy reaches 94.2\% on the held-out test set, with per-class F1 scores ranging from 0.91 (wireless microphone) to 0.97 (vacant). Decision latency from IQ capture to classification output averages 23~ms (median) with P95 at 31~ms, dominated by the spectrogram window; inference requires 2.1~ms on the H200 GPU.

\subsection{Super-Resolution Quality Assessment}
The SR pipeline substantially exceeds bicubic upscaling in PSNR/SSIM and improves downstream task performance measured by maritime object detection (YOLOv8 fine-tuned on SeaDronesSee~\cite{ref:seadronesee2022}). Processing sustains 30~fps output from 15~fps input with frame-to-frame latency of 45~ms (P95: 62~ms).

\subsection{RAN Co-tenancy Validation}
Over 24-hour continuous operation with spectrum sensing running at 1~Hz and super-resolution at full 30~fps output, we recorded zero timing violations across 172.8 million slots. CPU core isolation and GPU stream prioritization proved essential; removing either introduced sporadic violations.

\subsection{End-to-End System Evaluation}
Across five trial runs simulating intermittent PAWS connectivity, the system maintained video connectivity throughout outage periods. When connectivity was restored, database grants confirmed sensing recommendations in 97\% of cases; discrepancies involved channels reserved for wireless microphones in the database but without detectable transmissions during sensing.

\section{Discussion}\label{sec:discussion}
\subsection{Regulatory Pathway for SIDS Adoption}
For jurisdictions with PAWS infrastructure, the system functions as a standard database-governed TVWS device with enhanced resilience via sensing-based fallback. For jurisdictions without PAWS, we propose a pathway based on emergency waiver provisions paired with comprehensive audit logging for post-event review.

We are engaging with the Barbados Telecommunications Unit on a formal trial agreement enabling sensing-first operation during the CaribeWave~2026 Caribbean Tsunami Exercise (March 2026).

\subsection{Limitations and Future Work}
Our sensing model is trained primarily on Barbados spectrum data and may require retraining for other SIDS environments; transfer learning could reduce this burden but remains future work. Extending from a single-vessel deployment to cooperative multi-node sensing is planned for the CaribeWave~2026 trial. Additional validation is needed for real disaster RF conditions, including damaged transmitters and spurious emissions.

\section{Conclusion}\label{sec:conclusion}
We presented \textit{SIDSense}, an edge AI framework that enables disaster-resilient TVWS operation through CNN-based spectrum sensing, compliance-gated mode control, and GPU-aware RAN co-tenancy. Barbados deployments demonstrate 94.2\% sensing accuracy, zero missed L1 deadlines under concurrent AI workloads, and sustained video connectivity during simulated PAWS outages.

\section*{Acknowledgments}
This work was conducted under the AI-RAN Alliance Working Group 3 (AI-on-RAN) framework. We thank the Government of Barbados Ministry of Innovation, Industry, Science and Technology (MIST), and Barbados Telecommunications Unit for regulatory guidance and spectrum access authorization.


\begin{thebibliography}{99}
\bibitem{ref:undrr2022}
UNDRR, ``Global Assessment Report on Disaster Risk Reduction 2022: Our World at Risk,'' United Nations Office for Disaster Risk Reduction, Geneva, 2022.

\bibitem{ref:itu2017dominica}
ITU, ``Restoring Telecommunications in Dominica after Hurricane Maria,'' \emph{ITU News}, Nov. 2017.

\bibitem{ref:mwangama2015africon}
P.~Mwangama \emph{et al.}, ``TV White Spaces in Africa: From Research to Real-World Deployments,'' in \emph{Proc. IEEE AFRICON}, 2015.

\bibitem{ref:microsoft2019airband}
Microsoft, ``Airband Initiative: Expanding Broadband Access,'' Microsoft Corporate Report, 2019.

\bibitem{ref:rfc7545}
V.~Chen \emph{et al.}, ``Protocol to Access White-Space (PAWS) Databases,'' RFC 7545, IETF, May 2015.

\bibitem{ref:iana_paws}
IANA, ``PAWS Ruleset ID Registry.'' [Online]. Available: \url{https://www.iana.org/assignments/paws-ruleset-ids/}

\bibitem{ref:fcc2008}
FCC, ``Second Report and Order and Memorandum Opinion and Order,'' ET Docket No. 04-186, Nov. 2008.

\bibitem{ref:kawade2010dyspan}
S.~Kawade and M.~Nekovee, ``Can Cognitive Radio Access to TV White Spaces Support Future Home Networks?'' in \emph{Proc. IEEE DySPAN}, 2010.

\bibitem{ref:stevenson80222}
C.~R. Stevenson \emph{et al.}, ``IEEE 802.22: The First Cognitive Radio Wireless Regional Area Network Standard,'' \emph{IEEE Communications Magazine}, vol.~47, no.~1, pp.~130--138, Jan. 2009.

\bibitem{ref:nekovee2010survey}
M.~Nekovee, ``A Survey of Cognitive Radio Access to TV White Spaces,'' \emph{International Journal of Digital Multimedia Broadcasting}, 2010.

\bibitem{ref:ghosh2012broadcast}
A.~Ghosh \emph{et al.}, ``Coexistence of TV White Space Devices with Digital Television,'' \emph{IEEE Transactions on Broadcasting}, vol.~58, no.~3, 2012.

\bibitem{ref:murty2012senseless}
R.~Murty \emph{et al.}, ``SenseLess: A Database-Driven White Spaces Network,'' \emph{IEEE Transactions on Mobile Computing}, vol.~11, no.~2, 2012.

\bibitem{ref:zennaro2013africon}
M.~Zennaro \emph{et al.}, ``On the Design of a TV White Space Wireless Network for Rural Areas,'' in \emph{Proc. IEEE AFRICON}, 2013.

\bibitem{ref:lysko2013africon}
A.~Lysko \emph{et al.}, ``First Large TV White Spaces Trial in South Africa: A Brief Overview,'' in \emph{Proc. IEEE AFRICON}, 2013.

\bibitem{ref:falade2013commag}
A.~Falade \emph{et al.}, ``TV White Space Broadband for Rural Areas,'' \emph{IEEE Communications Magazine}, vol.~51, no.~7, 2013.

\bibitem{ref:johnson2017dev}
D.~Johnson and G.~van Stam, ``The Good, the Bad, and the Real: Reflections on Ten Years of Community Networks in the Global South,'' in \emph{Proc. ACM DEV}, 2017.

\bibitem{ref:haykin2005cr}
S.~Haykin, ``Cognitive Radio: Brain-Empowered Wireless Communications,'' \emph{IEEE Journal on Selected Areas in Communications}, vol.~23, no.~2, 2005.

\bibitem{ref:yucek2009survey}
T.~Y\"ucek and H.~Arslan, ``A Survey of Spectrum Sensing Algorithms for Cognitive Radio Applications,'' \emph{IEEE Communications Surveys \& Tutorials}, vol.~11, no.~1, 2009.

\bibitem{ref:zheng2020chinacom}
S.~Zheng \emph{et al.}, ``Spectrum Sensing Based on Deep Learning Classification for Cognitive Radios,'' \emph{China Communications}, vol.~17, no.~2, 2020.

\bibitem{ref:oshea2018jsps}
T.~J. O'Shea \emph{et al.}, ``Over-the-Air Deep Learning Based Radio Signal Classification,'' \emph{IEEE Journal of Selected Topics in Signal Processing}, vol.~12, no.~1, 2018.

\bibitem{ref:restuccia2019mobihoc}
F.~Restuccia \emph{et al.}, ``DeepRadioID: Real-Time Channel-Resilient Optimization of Deep Learning-based Radio Fingerprinting Algorithms,'' in \emph{Proc. ACM MobiHoc}, 2019.

\bibitem{ref:bao2010commag}
J.~Bao \emph{et al.}, ``Rapidly Deployable Emergency Networks,'' \emph{IEEE Communications Magazine}, vol.~48, no.~8, 2010.

\bibitem{ref:liu2013icist}
Y.~Liu \emph{et al.}, ``Emergency Communication Based on Wireless Mesh Networks,'' in \emph{Proc. IEEE ICIST}, 2013.

\bibitem{ref:saarnisaari2019aess}
H.~Saarnisaari \emph{et al.}, ``Satellite-Terrestrial Network Integration for Emergency Communications,'' \emph{IEEE Aerospace and Electronic Systems Magazine}, vol.~34, no.~10, 2019.

\bibitem{ref:pratas2012vtc}
N.~Pratas \emph{et al.}, ``Cooperative Spectrum Sensing for Disaster Relief Scenarios,'' in \emph{Proc. IEEE VTC Spring}, 2012.

\bibitem{ref:iturp1546}
ITU-R, ``Recommendation P.1546-6: Method for Point-to-Area Predictions for Terrestrial Services in the Frequency Range 30 MHz to 4000 MHz,'' 2019.

\bibitem{ref:wang2021iotj}
J.~Wang \emph{et al.}, ``Maritime Communications: A Survey on Enabling Technologies, Opportunities, and Challenges,'' \emph{IEEE Internet of Things Journal}, vol.~8, no.~19, 2021.

\bibitem{ref:polese2023oran}
M.~Polese \emph{et al.}, ``Understanding O-RAN: Architecture, Interfaces, Algorithms, Security, and Research Challenges,'' \emph{IEEE Communications Surveys \& Tutorials}, vol.~25, no.~2, 2023.

\bibitem{ref:airan2024report}
AI-RAN Alliance, ``Technical Report: AI-Native RAN Architecture,'' 2024.

\bibitem{ref:bonati2023coloran}
L.~Bonati \emph{et al.}, ``ColO-RAN: Developing Machine Learning-based xApps for Open RAN Closed-loop Control on Programmable Experimental Platforms,'' \emph{IEEE Transactions on Mobile Computing}, vol.~22, no.~10, 2023.

\bibitem{ref:tehrani2023cusense}
P.~Tehrani \emph{et al.}, ``cuSense: GPU-Accelerated Spectrum Sensing for Cognitive Radio,'' in \emph{Proc. IEEE INFOCOM}, 2023.

\bibitem{ref:riyaz2018commag}
S.~Riyaz \emph{et al.}, ``Deep Learning Convolutional Neural Networks for Radio Identification,'' \emph{IEEE Communications Magazine}, vol.~56, no.~9, 2018.

\bibitem{ref:realesrgan2021}
X.~Wang \emph{et al.}, ``Real-ESRGAN: Training Real-World Blind Super-Resolution with Pure Synthetic Data,'' in \emph{Proc. ICCV Workshops}, 2021.

\bibitem{ref:rife2022}
Z.~Huang \emph{et al.}, ``RIFE: Real-Time Intermediate Flow Estimation for Video Frame Interpolation,'' in \emph{Proc. ECCV}, 2022.

\bibitem{ref:seadronesee2022}
L.~Varga \emph{et al.}, ``SeaDronesSee: A Maritime Benchmark for Detecting Humans in Open Water,'' in \emph{Proc. WACV}, 2022.
\end{thebibliography}
\end{document}